\documentclass[12pt]{article}

\usepackage{amssymb}
\usepackage{amsmath}
\usepackage{graphicx}
\usepackage[hyperindex]{hyperref}

\newcommand{\M}{\mathcal{M}}

\newcommand{\Scal}{\Sigma}
\newcommand{\Rcal}{\mathcal{R}}

\newcommand{\past}{\mathbf{past}}
\newcommand{\fut}{\mathbf{future}}

\newcommand{\expec}[1]{\left<#1\right>}

\begin{document}

\begin{center}
{\normalsize \hfill IMPERIAL-TP-06-SZ-06}\\

\vspace{60pt}
{ \Large \bf Counting entropy in causal set quantum gravity}

\vspace{30pt}

{\sl D. Rideout} and
{\sl S. Zohren}

\vspace{30pt}
{\footnotesize

Blackett Laboratory, \\
Imperial College,\\
London SW7 2AZ, United Kingdom.\\
\vspace{10pt}
Email: \\
{\sl d.rideout@imperial.ac.uk}, {\sl stefan.zohren@imperial.ac.uk}\\

}
\vspace{50pt}

\begin{abstract}
The finiteness of black hole entropy suggest that spacetime is fundamentally discrete, and hints at an underlying relationship between geometry and "information". The foundation of this relationship is yet to be uncovered, but should manifest itself in a theory of quantum gravity. We review recent attempts to define a microscopic measure for black hole entropy and for the maximum entropy of spherically symmetric spacelike regions, within the causal set approach to quantum gravity.
\end{abstract}

\vspace{160pt}

{\footnotesize Talk given by S. Zohren at the Eleventh Marcel Grossmann Meeting on General Relativity at the Freie U. Berlin, July 23 - 29, 2006.}

\end{center}
\newpage

\section{Introduction}\label{intro}
The various entropy bounds that exist in the literature (see for a review \cite{Bousso:2002ju}) suggest that an underlying theory
of quantum gravity should predict these bounds from a counting of microstates and should clarify which are the fundamental degrees of freedom one is actually counting. This verification of the thermodynamic laws is an important consistency check for any approach to quantum gravity.

In what follows we review an earlier work by Dou and Sorkin\cite{Dou:2003af} defining a microscopic measure for black hole entropy together with our recent proposal\cite{Rideout:2006zt} for measuring the maximum entropy contained in a spherically symmetric spacelike region, within the causal set approach to quantum gravity.

\section{Causal set quantum gravity}
Causal set theory is an approach to fundamentally discrete quantum gravity (see for a recent review \cite{Henson:2006kf}).
Besides taking fundamental discreteness as a first principle, the primacy of causal structure 
is the main observation underlying causal sets. 

Mathematically a causal set is a locally finite partially ordered set, or in other words a set $C$ endowed with a binary relation `precedes' $\prec$, which satisfies: 
$(i)$ \textit{transitivity}: if $x\prec y$ and $y\prec z$ then $x\prec z$, $(ii)$ \textit{irreflexivity}: $x \not\prec x$, $(iii)$ \textit{local finiteness}: for any pair of elements $x$ and $z$ of $C$, the set of elements lying between $x$ and $z$ is finite, $|\{y | x \prec y \prec z \}|<\infty$. Some useful definitions are the past of an element $\past(x)=\{y\in C|y\prec x\}$ and its future $\fut(x)=\{y\in C|x\prec y\}$. Further, a relation $x\prec y$ is called a link iff $\fut(x)\cap\past(y)=\{x,y\}$. Elements of the causal set whose future (past) is empty
are called maximal (minimal). 

The hypothesis of causal set theory is that spacetime at short scales such as the Planck length is fundamentally discrete, and is better described by a causal set than a differentiable manifold. The notion of continuum Lorentzian spacetime $\M$ at larger scales is recovered as an approximation of the causal set. 
This occurs when the causal set can be faithfully embedded into $\M$, where faithfully means that the embedding respects not only the causal relations, but also
a correspondence between cardinality and spacetime volume.

\begin{figure}[t]
\begin{center}
\includegraphics[width=5.5in]{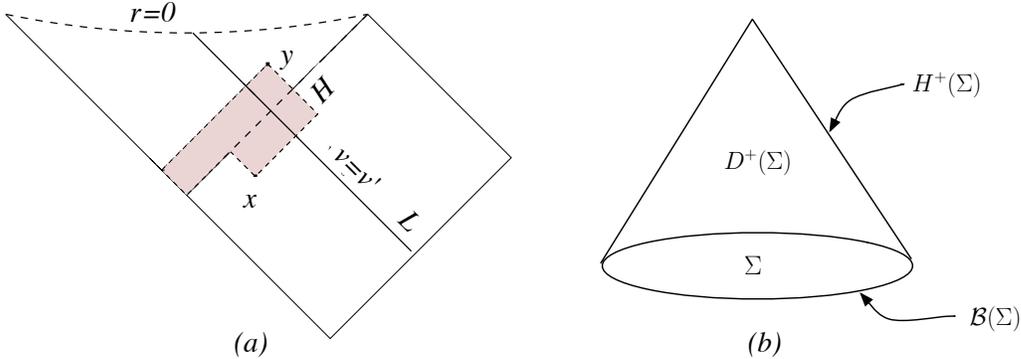}
\caption{(a) Schwarzschild spacetime
and null hypersurface $L$. (b) Spherically symmetric spacelike region $\Scal$, its future domain of dependence $D^+(\Scal)$ and future Cauchy horizon $H^+(\Scal)$.}
\label{default}
\end{center}
\end{figure}

\section{Black hole entropy}

In an earlier work, Dou and Sorkin considered the four-dimensional Schwarzschild black hole in its dimensionally reduced form, 
$ds^2=-4a^3/r\,e^{-r/a}dudv,$ where $a$ is the Schwarzschild radius of the black hole and $u$ and $v$ are the Kruskal coordinates \cite{Dou:2003af}.  Assuming that this spacetime arises as an approximation to a causal set which can be faithfully embedded into it, 
they propose to count the number of causal links
from causal set elements $x\in\Rcal_1=J^-(H)\cap J^-(L)$ to elements $y\in\Rcal_2=J^+(H)\cap J^+(L)$ (see fig. (a)). 
The motivation for counting links comes from regarding the black hole entropy
as arising from quantum entanglement across the horizon $H$ evaluated at a null hypersurface $L$, and noting that the links are effectively
irreducible elements of potential information flow in a causal set.
The number of such links is given by
$\expec{n}=\int_{\Rcal_1}\int_{\Rcal_2} e^{-V(x,y)} dV_x  dV_y$, where
$V(x,y)$ denotes the volume of $J^+(x)\cap J^-(y)$. 
(The dimensional reduction is necessary to make feasible the computation of such regions.)
To suppress certain
unphysical nonlocal links one further has to impose that the elements $y$ are
minimal in $J^+(H)$. 
Evaluating the above integral at scales much larger than the discreteness scale then yields $\expec{n}=\pi^2/6+ \cdots$ (where the $\cdots$ represent higher order terms in the ratio of the discreteness scale to the macroscopic scale $a$).
Unfortunately, when one considers the angular dimensions, it now seems clear that the expected number of links will diverge, essentially because the intersection of the future light cone of a candidate element $x$ with $H$ has an infinite extent.  However, it seems likely that a minor variation, such as counting triples of elements rather than pairs, will lead to a convergent integral in the full four-dimensional case.


\section{The spherical entropy bound}
We now 
discuss our recently proposed microscopic evidence for the spherical entropy bound arising from causal set theory. Susskind's spherical entropy bound\cite{Susskind:1994vu} states that the entropy of the matter content of a spherically symmetric spacelike region $\Scal$ (of finite volume) is bounded by a quarter of the area of the boundary of $\Scal$ in Planck units, $S\leqslant A/(4l_P^2)$, where $l_P$ is the Planck length.

In the case of black holes the 
 counting of links is computationally difficult in the full four-dimensional geometry, because of the complicated causal structure in the angular coordinates.
For the simpler case of the spherically symmetric region 
$\Scal$ let us now propose the following 
measure of entropy. Note that the entropy of the matter contained in $\Scal$
must eventually ``flow out'' of the region by passing over the boundary of its future domain of dependence $D^+(\Scal)$, the future Cauchy horizon $H^+(\Scal)$ (see fig. (b)).  But because spacetime is fundamentally discrete, the amount of such entropy flux is bounded above by the number of discrete elements comprising this boundary. These elements can be seen as just the maximal elements of the causal set faithfully embedded into the future domain of dependence $D^+(\Scal)$. 
This is similar to the case of the black hole, where the links started at the elements $x$ which were maximal in $\Rcal_1$ (by definition of being linked to $y$).
Hence we define the \textit{maximal entropy} contained in $\Scal$ as the number of maximal elements in $D^+(\Scal)$,
$S_{max}=\expec{n}=\int_{D^+(\Sigma)} e^{-V(x)} dV_x,$
where $V(x)$ is the volume of $\fut(x)\cap D^+(\Scal)$. 
The claim is that if the fundamental discreteness scale is fixed at a dimension-dependent value this proposal leads to Susskind's 
spherical entropy bound in the continuum approximation, $S_{max}\!=\!A/(4l_P^2)$, where $A$ is the area of the boundary of $\Scal$.

For the case where $\Scal$ is a three dimensional-ball in four-dimensional Minkowski spacetime, $\expec{n}$ can be evaluated analytically yielding,
at scales much larger than the discreteness scale,
$\expec{n}=\sqrt{6}\,A/(4l_P^2)+ \cdots$.
This shows that indeed the result is proportional to the area of the boundary of $\Scal$. 
If we fix the fundamental discreteness scale to $l_f=\sqrt[4]{6}\, l_P$, we arrive at the desired result $S_{max}=A/(4l_P^2)$. Further, we could numerically show that one obtains the same result 
in the case of different spherically symmetric spacelike regions in four-dimensional Minkowski spacetime as well as for different dimensions, where the value of the fundamental discreteness scale changed with the dimension. Work in progress indicates that this result is also true in the case of conformally flat Friedmann-Robertson-Walker spacetime.

\section*{Acknowledgments}
The authors acknowledge support by
the European Network on
Random Geometry, ENRAGE (MRTN-CT-2004-005616).
Further, we would like to
thank F.\ Dowker for enjoyable discussions, comments, and critical proof
reading of the manuscript.

\providecommand{\href}[2]{#2}\begingroup\raggedright

\end{document}